\documentclass[letterpaper,english,reprint, aps, 10pt, superscriptaddress]{revtex4-1}
\usepackage[T1]{fontenc}
\usepackage[latin9]{inputenc}
\usepackage{color}
\usepackage{babel}
\usepackage{array}
\usepackage{amsmath}
\usepackage{amssymb}
\usepackage{graphicx}
\usepackage[unicode=true,pdfusetitle,
 bookmarks=true,bookmarksnumbered=false,bookmarksopen=false,
 breaklinks=true,pdfborder={0 0 1},backref=false,colorlinks=true]
 {hyperref}
\hypersetup{
 urlcolor=blue, linkcolor=blue, citecolor=blue}

\makeatletter

\pdfpageheight\paperheight
\pdfpagewidth\paperwidth

\providecommand{\tabularnewline}{\\}

\makeatother

\begin{document}
\preprint{APS/123-QED}
\title{Application of machine learning potentials to predict grain boundary
properties\\
in fcc elemental metals}
\author{Takayuki \surname{Nishiyama}}
\email{nishiyama.takayuki.84e@st.kyoto-u.ac.jp}

\affiliation{Department of Materials Science and Engineering, Kyoto University,
Kyoto 606-8501, Japan}
\author{Atsuto \surname{Seko}}
\email{seko@cms.mtl.kyoto-u.ac.jp}

\affiliation{Department of Materials Science and Engineering, Kyoto University,
Kyoto 606-8501, Japan}
\author{Isao \surname{Tanaka}}
\affiliation{Department of Materials Science and Engineering, Kyoto University,
Kyoto 606-8501, Japan}
\affiliation{Center for Elements Strategy Initiative for Structure Materials (ESISM),
Kyoto University, Kyoto 606-8501, Japan}
\affiliation{Nanostructures Research Laboratory, Japan Fine Ceramics Center, Nagoya
456-8587, Japan}
\date{\today}
\begin{abstract}
Accurate interatomic potentials are in high demand for large-scale
atomistic simulations of materials that are prohibitively expensive
by density functional theory (DFT) calculation. In this study, we
apply machine learning potentials in a recently constructed repository
to the prediction of the grain boundary energy in face-centered-cubic
elemental metals, i.e., Ag, Al, Au, Cu, Pd, and Pt. The systematic
application of machine learning potentials shows that they enable
us to predict grain boundary structures and their energies accurately.
The grain boundary energies predicted by the MLPs are in agreement
with those calculated by DFT, although no grain boundary structures
were included in training datasets of the present MLPs.
\end{abstract}
\maketitle

\section{Introduction}

Grain boundaries are interfaces between differently oriented crystals
of the same phase \citep{Sutton1995}. The microstructures of grain
boundaries can affect various properties of polycrystalline materials,
including mechanical, thermal, and electrical properties \citep{Bishop1951,Kroner1961,Mishin2010,Greer2011}.
Thus, an attractive topic in materials science has been to establish
the relationship between the properties of crystalline materials and
grain boundary structures. Many theoretical studies have been made
to cover a broad range of grain boundary structures and their excessive
energies. Early fundamental studies employed pair potentials, such
as the Lennard\textendash Jones and Morse forms, to investigate the
generic properties of grain boundaries such as the presence of cusps
in a map of the rotation angle and the grain boundary energy \citep{Hasson1971,Wolf1989,Merkle1990}.
Empirical interatomic potentials such as the Finnis\textendash Sinclair
(FS) potentials \citep{Finnis1984} and embedded atom method (EAM)
\citep{Daw1984} potentials have been widely used to investigate symmetric
and asymmetric grain boundaries of metallic materials. Quantitative
predictions are becoming possible \citep{Cahn2006,VonAlfthan2006,Brown2007,Dao2007,Tschopp2007,Olmsted2009,Sangid2010,Tschopp2012,Tschopp2015,Kiyohara2016},
and strong correlations between theoretical and experimental grain
boundary energies have been shown, especially for grain boundaries
in elemental Al and Ni, which exhibit low grain boundary energies
\citep{Rohrer2010,Holm2011}. However, the prediction error in the
grain boundary energy may be significant in grain boundaries showing
higher grain boundary energies. This error originates from the fact
that their microscopic grain boundary structures differ from the atomic
environment used to estimate interatomic potentials.

Density functional theory (DFT) calculation \citep{Hohenberg1964,Kohn1965}
is an alternative way to predict grain boundary properties accurately.
However, DFT calculation is practically impossible to apply to large-scale
models of grain boundaries owing to its computational cost. Therefore,
interatomic potentials that enable us to predict grain boundary properties
accurately have been in high demand. Over the last decade, many groups
have proposed frameworks to develop machine learning potentials (MLPs)
based on extensive datasets generated by DFT calculation \citep{Lorenz2004210,Behler2007,Bartok2010,Behler2011,Han2018,258c531ae5de4f5699e2eec2de51c84f,PhysRevB.96.014112,PhysRevB.90.104108,PhysRevX.8.041048,PhysRevLett.114.096405,PhysRevB.95.214302,PhysRevB.90.024101,PhysRevB.92.054113,PhysRevMaterials.1.063801,Thompson2015316,Wood2018,PhysRevMaterials.1.043603,doi-10.1137-15M1054183,PhysRevLett.120.156001,Podryabinkin2019,GUBAEV2019148,doi:10.1063/1.5126336}.
The MLPs significantly improve the accuracy and transferability of
interatomic potentials. Also, MLPs themselves are becoming available,
such as those in \textsc{Machine Learning Potential Repository}\citep{Seko2020}
developed by one author of this paper.

In this paper, we demonstrate the predictive power of MLPs in the
MLP repository for grain boundary properties. We systematically evaluate
the structures and excessive energies of $\langle100\rangle$ symmetric
tilt grain boundaries (STGBs), $\langle110\rangle$ STGBs, and $\langle100\rangle$
pure-twist grain boundaries in the face-centered-cubic (fcc) elemental
metals of Ag, Al, Au, Cu, Pd, and Pt. They are compared with those
obtained from EAM potentials and DFT calculations. The MLP repository
contains a set of Pareto optimal MLPs with different trade-offs between
accuracy and computational efficiency; hence, we carefully determine
appropriate MLPs to predict grain boundary properties.

\section{Methodology}

\subsection{Modeling and structure optimization of grain boundaries}

Macroscopic structures of grain boundaries are characterized by five
geometrical degrees of freedom. We choose three variables to specify
the direction of the rotation axis and the rotation angle, which describe
the misorientation between crystal lattices, and two variables to
specify the direction of the boundary plane normal \citep{Sutton1995}.
For a given set of macroscopic variables, the microscopic structure
is associated with three degrees of freedom regarding rigid body displacements:
two components parallel to the boundary plane and one component normal
to the plane. Hence, the globally optimal microscopic structure for
a given set of macroscopic variables is achieved by optimizing the
three microscopic variables in terms of potential energy.

In this study, we investigate only STGBs and pure-twist grain boundaries.
The periodicity of an STGB is identified from the orthogonal projection
of its coincident site lattice (CSL) to its boundary plane. Also,
the periodicity of a pure-twist grain boundary is given by the orthogonal
projection of its displacement shift complete (DSC) lattice to its
boundary plane. Therefore, we restrict the ranges of the two in-plane
microscopic variables to a domain defined by the periodicity of the
grain boundaries.

We explore the globally optimal microscopic structure for a set of
macroscopic variables using a multi-start method. The multi-start
method involves local structure optimizations for a given set of initial
structures and regards the structure with the lowest energy among
the converged final structures as the globally optimal structure.
We use the conjugate gradient method implemented in the \textsc{lammps}
code \citep{Plimpton1995} for the local structure optimizations.
Initial microscopic structures are introduced from a 4 $\times$ 4
grid for the two in-plane components and a sequence for the component
normal is introduced to the boundary plane. For each initial microscopic
structure, a calculation model is generated using \textsc{pymatgen}
\citep{Ong2013}. This model contains two parallel boundaries perpendicular
to the c-axis of the model, separated by fcc layers corresponding
to four repetitions of a cell of the CSL. However, the local structure
optimization starting from some of the initial microscopic structures
fails to converge when using both the MLPs and the EAM potentials,
as shown in the next section. These structures are ignored in finding
the globally optimal microscopic structure. Note that the optimization
of the microscopic structure is performed in the whole domain here,
although it is more efficient to restrict the domain to its symmetrically
nonequivalent domain.

\subsection{Machine learning potentials}

We employ MLPs in \textsc{Machine Learning Potential Repository}
\citep{Seko2020} developed by one author of this paper to obtain
the globally optimal microscopic structures of STGBs and pure-twist
grain boundaries. In the repository, a set of Pareto optimal MLPs
with different trade-offs between accuracy and computational efficiency
is available, from which one can choose an appropriate MLP in accordance
with the target and purpose. Potential energy models of the MLPs are
either a polynomial model of Gaussian-type pairwise structural features
or a polynomial model of polynomial invariants for the O(3) group,
which are derived by a group-theoretical approach \citep{Seko2019}.

The Pareto optimal MLPs in the repository have been developed using
a dataset generated from structure generators. For Ag, Al, Au, and
Cu, we adopt the Pareto optimal MLPs developed from a structure generator
set composed of the fcc, body-centered-cubic (bcc), hexagonal-close-packed
(hcp), simple cubic (sc), $ \omega $, and $ \beta $ tin structures.
The dataset is composed of 3,000 structures constructed by introducing
random lattice expansion, random lattice distortion, and random atomic
displacements into a supercell of the equilibrium structure for one
of the structure generators. For Pd and Pt, we employ another set
of 82 prototype structures as the structure generator set because
the dataset derived from the six structure generators is not available
in the repository. The dataset consists of 10,000 structures generated
by the same procedure as above. For all structures in the dataset,
DFT calculations were performed using the plane-wave-basis projector
augmented wave method \citep{Blochl1994} within the Perdew\textendash Burke\textendash Ernzerhof
exchange-correlation functional \citep{Perdew1996} as implemented
in the \textsc{VASP} code \citep{Kresse1993,Kresse1996,Kresse1999}.
Note that the datasets contain no structures generated from grain
boundary models.

\section{Results and discussion}

First, we systematically calculate the grain boundary energies of
five grain boundaries using the whole set of Pareto optimal MLPs for
each elemental metal. They are the $\Sigma 5$ $\langle100\rangle$
STGB (at 53.1 degrees), the $\Sigma 3$ $\langle110\rangle$ STGB
(at 70.5 degrees), the $\Sigma 3$ $\langle110\rangle$ STGB (at 109.5
degrees), the $\Sigma 9$ $\langle110\rangle$ STGB (at 38.9 degrees),
and the $\Sigma 5$ $\langle100\rangle$ pure-twist grain boundary
(at 36.9 degrees), the calculation models for which can be represented
by a small number of atoms. Then, we find an accurate MLP requiring
only a reasonable computational time to investigate the whole set
of grain boundaries.

\begin{figure*}
\begin{tabular}{c>{\centering}p{0.28\linewidth}>{\centering}p{0.28\linewidth}>{\centering}p{0.28\linewidth}}
 & $\langle100\rangle$ STGB & $\langle110\rangle$ STGB & $\langle100\rangle$ pure-twist grain boundary\tabularnewline
Ag & %
\begin{tabular}{c}
\includegraphics[width=1\linewidth]{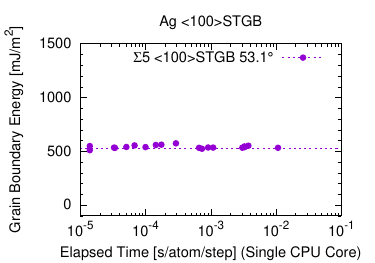}\tabularnewline
\end{tabular} & %
\begin{tabular}{c}
\includegraphics[width=1\linewidth]{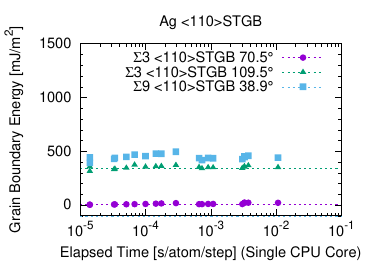}\tabularnewline
\end{tabular} & %
\begin{tabular}{c}
\includegraphics[width=1\linewidth]{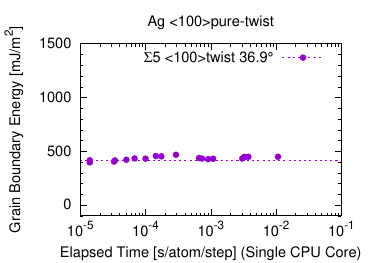}\tabularnewline
\end{tabular}\tabularnewline
Al & %
\begin{tabular}{c}
\includegraphics[width=1\linewidth]{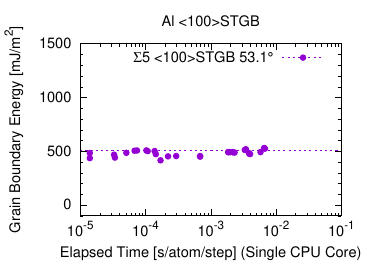}\tabularnewline
\end{tabular} & %
\begin{tabular}{c}
\includegraphics[width=1\linewidth]{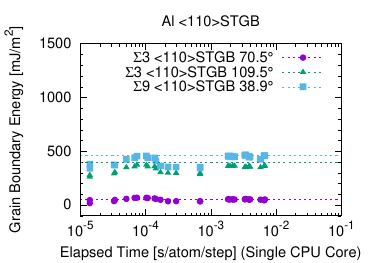}\tabularnewline
\end{tabular} & %
\begin{tabular}{c}
\includegraphics[width=1\linewidth]{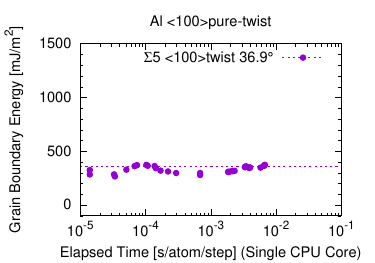}\tabularnewline
\end{tabular}\tabularnewline
Au & %
\begin{tabular}{c}
\includegraphics[width=1\linewidth]{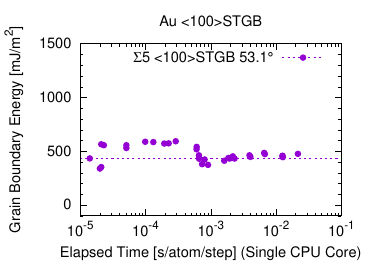}\tabularnewline
\end{tabular} & %
\begin{tabular}{c}
\includegraphics[width=1\linewidth]{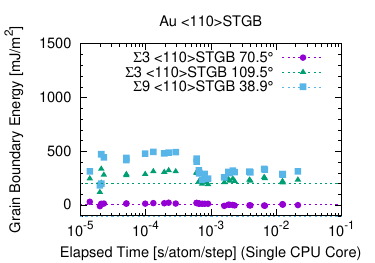}\tabularnewline
\end{tabular} & %
\begin{tabular}{c}
\includegraphics[width=1\linewidth]{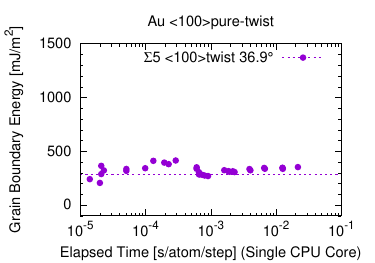}\tabularnewline
\end{tabular}\tabularnewline
Cu & %
\begin{tabular}{c}
\includegraphics[width=1\linewidth]{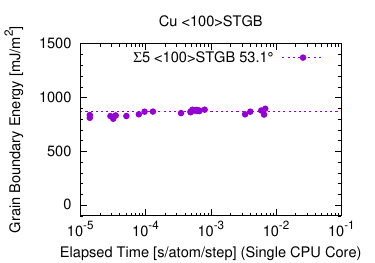}\tabularnewline
\end{tabular} & %
\begin{tabular}{c}
\includegraphics[width=1\linewidth]{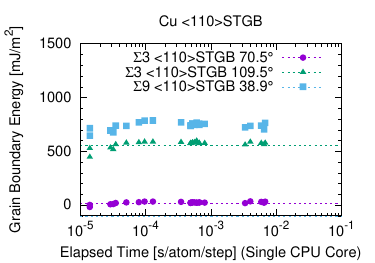}\tabularnewline
\end{tabular} & %
\begin{tabular}{c}
\includegraphics[width=1\linewidth]{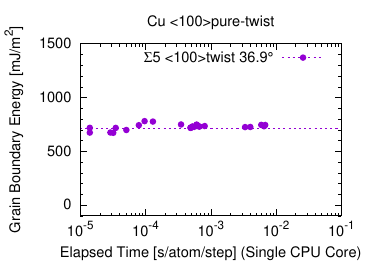}\tabularnewline
\end{tabular}\tabularnewline
Pd & %
\begin{tabular}{c}
\includegraphics[width=1\linewidth]{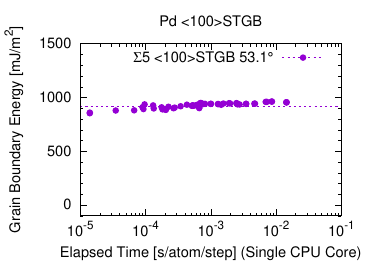}\tabularnewline
\end{tabular} & %
\begin{tabular}{c}
\includegraphics[width=1\linewidth]{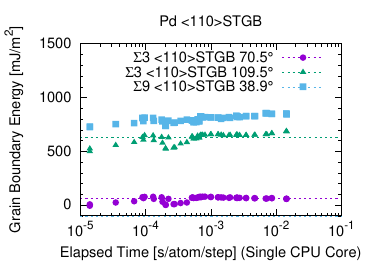}\tabularnewline
\end{tabular} & %
\begin{tabular}{c}
\includegraphics[width=1\linewidth]{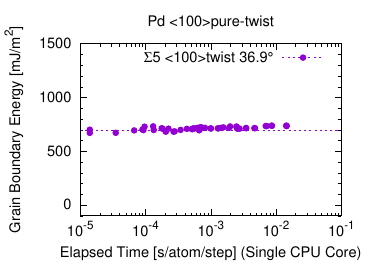}\tabularnewline
\end{tabular}\tabularnewline
Pt & %
\begin{tabular}{c}
\includegraphics[width=1\linewidth]{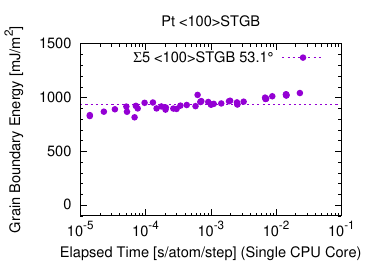}\tabularnewline
\end{tabular} & %
\begin{tabular}{c}
\includegraphics[width=1\linewidth]{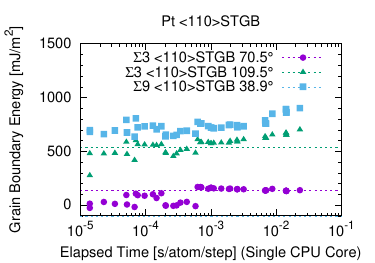}\tabularnewline
\end{tabular} & %
\begin{tabular}{c}
\includegraphics[width=1\linewidth]{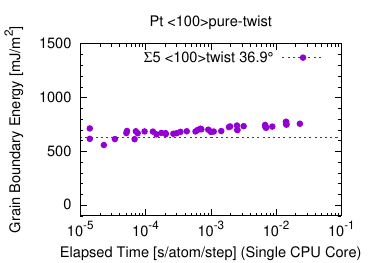}\tabularnewline
\end{tabular}\tabularnewline
\end{tabular}

\caption{\label{fig:transferability}Grain boundary energies of $\Sigma 5$
$\langle100\rangle$ STGB in 53.1 degrees, $\Sigma 3$ $\langle110\rangle$
STGB in 70.5 degrees, $\Sigma 3$ $\langle110\rangle$ STGB in 109.5
degrees, $\Sigma 9$ $\langle110\rangle$ STGB in 38.9 degrees, and
$\Sigma 5$ $\langle100\rangle$ pure-twist grain boundary in 36.9
degrees for elemental Ag, Al, Au, Cu, Pd, and Pt, predicted using
the Pareto optimal MLPs. The grain boundary energies computed by DFT
calculation are also shown by broken lines.}
\end{figure*}

Figure \ref{fig:transferability} shows the convergence behavior of
the grain boundary energy in terms of the computational time, obtained
using the whole set of Pareto optimal MLPs. The grain boundary energy
is identical to the lowest energy among the grain boundary energies
of the microscopic structures. The grain boundary energy of a microscopic
structure is measured from the energy of the equilibrium fcc structure.
The computational time corresponding to the model complexity of an
MLP is the elapsed time normalized by the number of atoms for a single
point calculation of the energy, the forces, and the stress tensors
\footnote{The computational time is estimated using a single core of Intel Xeon
E5-2695 v4 (2.10GHz).}. As can be seen in Figure \ref{fig:transferability}, the grain boundary
energy converges well in all of the elemental metals and grain boundaries.
Consequently, successive calculations for the whole set of grain boundaries
are performed using the MLP that requires the lowest computational
time among the MLPs showing convergence. 

We also examine the transferability of the MLPs to the prediction
of the grain boundary structures and energies because the datasets
used in developing the MLPs contain no grain boundary structures.
Therefore, we evaluate the grain boundary energies of the $\Sigma 3$
$\langle110\rangle$ STGB (at 70.5 degrees), the $\Sigma 3$ $\langle110\rangle$
STGB (at 109.5 degrees), the $\Sigma 9$ $\langle110\rangle$ STGB
(at 38.9 degrees), the $\Sigma 5$ $\langle100\rangle$ STGB (at 53.1
degrees), and the $\Sigma 5$ $\langle100\rangle$ pure-twist grain
boundary (at 36.9 degrees) by DFT calculation, and compare them with
those predicted using the MLPs. Figure \ref{fig:transferability}
shows the DFT values of the grain boundary energy only for the grain
boundary structures, DFT calculations for which converge successfully.
They are close to the grain boundary energies of the selected MLPs.
Therefore, the selected MLPs should have high predictive power for
grain boundary structures and their energies even though no grain
boundary structures were used to develop the MLPs.
\begin{table*}
\caption{\label{tab:Model-parameters-of}Model parameters of the MLPs used
to estimate the grain boundary structures and energies. The identification
of the feature type, the model type, and the polynomial orders can
be found in Ref. \citep{Seko2020}.}

\begin{ruledtabular}
\begin{tabular}{lcccccc}
 & Ag & Al & Au & Cu & Pd & Pt\tabularnewline
\hline 
MLP-ID & pair-44 & gtinv-336 & gtinv-111 & pair-23 & gtinv-722 & gtinv-533\tabularnewline
RMSE (energy) {[}meV/atom{]} & 2.2 & 0.8 & 0.7 & 2.2 & 6.3 & 12.9\tabularnewline
RMSE (force) {[}eV/$\text{Å}${]} & 0.010 & 0.008 & 0.012 & 0.013 & 0.097 & 0.172\tabularnewline
Time {[}ms/atom/step{]} \cite{Note1} & 0.05 & 1.85 & 0.66 & 0.04 & 0.52 & 0.63\tabularnewline
Number of coefficients & 815 & 1100 & 475 & 285 & 500 & 1595\tabularnewline
Feature type & Pair & Invariants & Invariants & Pair & Invariants & Invariants\tabularnewline
Cutoff radius {[}\AA{]} & 7.0 & 8.0 & 6.0 & 7.0 & 6.0 & 6.0\tabularnewline
Number of radial functions & 15 & 15 & 10 & 10 & 5 & 5\tabularnewline
Model type & 2 & 3 & 3 & 2 & 4 & 2\tabularnewline
Polynomial order (function $F$) & 3 & 3 & 3 & 3 & 2 & 2\tabularnewline
Polynomial order (invariants) & $-$ & 3 & 3 & $-$ & 3 & 3\tabularnewline
Spherical harmonics truncation $\{l_{\mathrm{max}}^{(2)}, l_{\mathrm{max}}^{(3)}\}$ & $-$ & {[}4, 4{]} & {[}4, 4{]} & $-$ & {[}4, 0{]} & {[}4, 2{]}\tabularnewline
\end{tabular}
\end{ruledtabular}

\end{table*}

Table \ref{tab:Model-parameters-of} lists the model parameters of
the selected MLPs. Fast MLPs are selected for Ag and Cu, while computationally
expensive MLPs are selected for the others. Table \ref{tab:Model-parameters-of}
also shows the prediction errors for the datasets used in developing
the MLPs. The MLPs for Pd and Pt show significant prediction errors,
which originate from the fact that the datasets contain many hypothetical
structures such as the graphite-type structure. Although the selected
MLPs exhibit significant prediction errors for such abnormal structures,
they show much smaller prediction errors for typical metallic structures,
including grain boundary structures, as shown above.

\begin{figure*}
\begin{tabular}{c>{\centering}p{0.28\linewidth}>{\centering}p{0.28\linewidth}>{\centering}p{0.28\linewidth}}
 & $\langle100\rangle$ STGB & $\langle110\rangle$ STGB & $\langle100\rangle$ pure-twist grain boundary\tabularnewline
Ag & %
\begin{tabular}{c}
\includegraphics[width=1\linewidth]{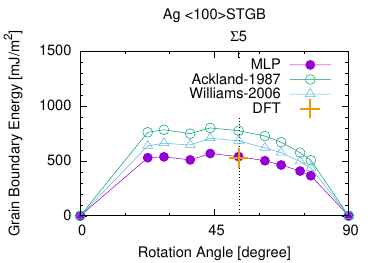}\tabularnewline
\end{tabular} & %
\begin{tabular}{c}
\includegraphics[width=1\linewidth]{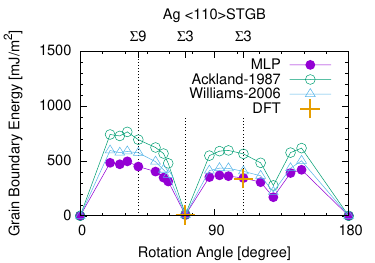}\tabularnewline
\end{tabular} & %
\begin{tabular}{c}
\includegraphics[width=1\linewidth]{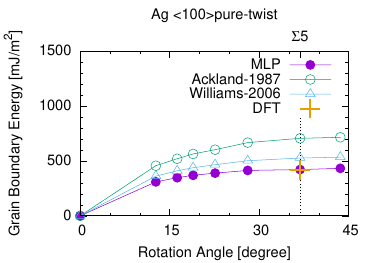}\tabularnewline
\end{tabular}\tabularnewline
Al & %
\begin{tabular}{c}
\includegraphics[width=1\linewidth]{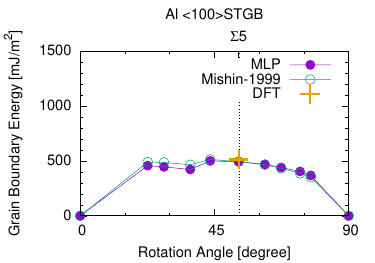}\tabularnewline
\end{tabular} & %
\begin{tabular}{c}
\includegraphics[width=1\linewidth]{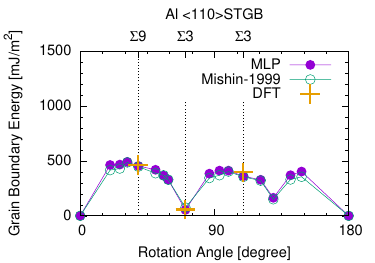}\tabularnewline
\end{tabular} & %
\begin{tabular}{c}
\includegraphics[width=1\linewidth]{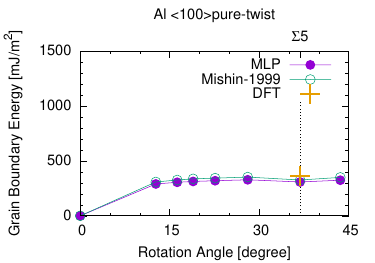}\tabularnewline
\end{tabular}\tabularnewline
Au & %
\begin{tabular}{c}
\includegraphics[width=1\linewidth]{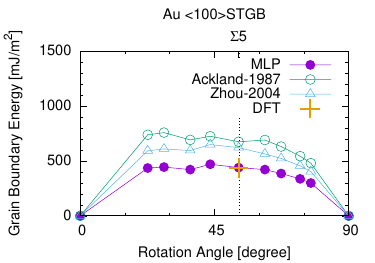}\tabularnewline
\end{tabular} & %
\begin{tabular}{c}
\includegraphics[width=1\linewidth]{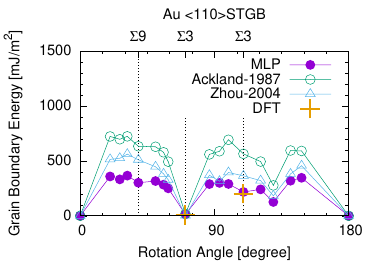}\tabularnewline
\end{tabular} & %
\begin{tabular}{c}
\includegraphics[width=1\linewidth]{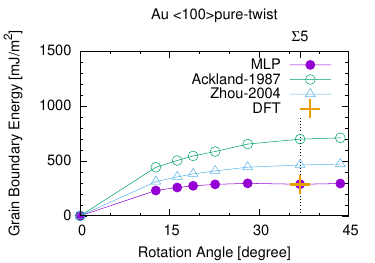}\tabularnewline
\end{tabular}\tabularnewline
Cu & %
\begin{tabular}{c}
\includegraphics[width=1\linewidth]{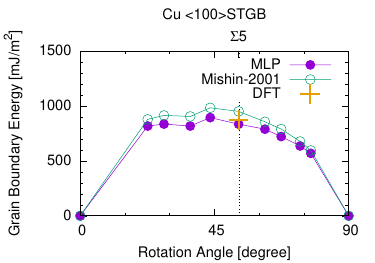}\tabularnewline
\end{tabular} & %
\begin{tabular}{c}
\includegraphics[width=1\linewidth]{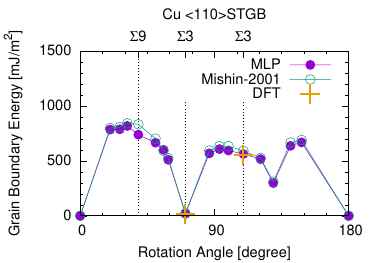}\tabularnewline
\end{tabular} & %
\begin{tabular}{c}
\includegraphics[width=1\linewidth]{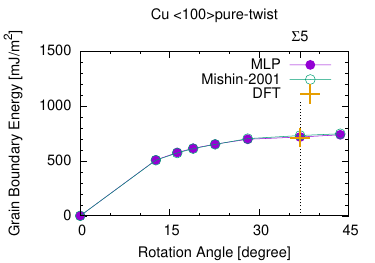}\tabularnewline
\end{tabular}\tabularnewline
Pd & %
\begin{tabular}{c}
\includegraphics[width=1\linewidth]{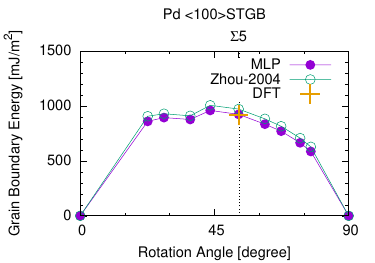}\tabularnewline
\end{tabular} & %
\begin{tabular}{c}
\includegraphics[width=1\linewidth]{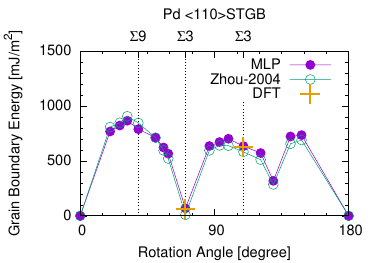}\tabularnewline
\end{tabular} & %
\begin{tabular}{c}
\includegraphics[width=1\linewidth]{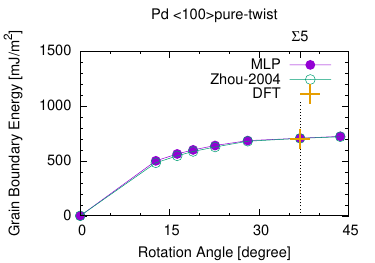}\tabularnewline
\end{tabular}\tabularnewline
Pt & %
\begin{tabular}{c}
\includegraphics[width=1\linewidth]{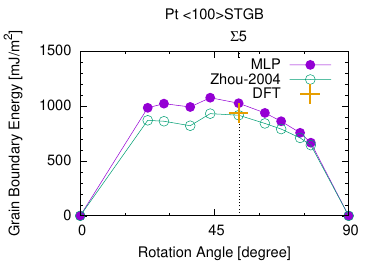}\tabularnewline
\end{tabular} & %
\begin{tabular}{c}
\includegraphics[width=1\linewidth]{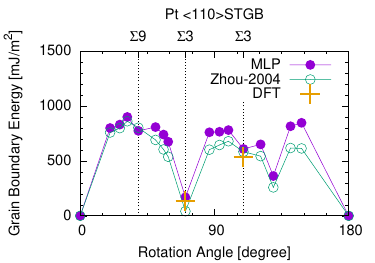}\tabularnewline
\end{tabular} & %
\begin{tabular}{c}
\includegraphics[width=1\linewidth]{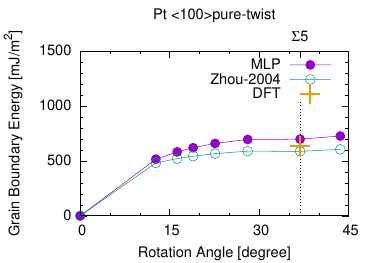}\tabularnewline
\end{tabular}\tabularnewline
\end{tabular}

\caption{\label{fig:gbenergy}Rotation angle dependence of the grain boundary
energy for $\langle100\rangle$ STGBs, $\langle110\rangle$ STGBs,
and $\langle100\rangle$ pure-twist grain boundaries for elemental
Ag, Al, Ag, Cu, Pd, and Pt, predicted using the MLPs. For comparison,
the grain boundary energies predicted using EAM potentials for Ag
\citep{Ackland1987,Williams2006}, Al \citep{Mishin1999}, Au \citep{Ackland1987,Zhou2004},
Cu \citep{Mishin2001}, Pd \citep{Zhou2004}, and Pt \citep{Zhou2004}
are shown by open symbols. The grain boundary energies computed by
DFT calculation are also shown by crosses.}
\end{figure*}

After confirming the transferability of the MLPs, we calculate the
energies of the grain boundary structures: $\langle100\rangle$ STGBs
($\Sigma5$, $\Sigma13$, $\Sigma17$, $\Sigma25$, $\Sigma29$, $\Sigma41$),
$\langle110\rangle$ STGBs ($\Sigma3$, $\Sigma9$, $\Sigma11$, $\Sigma17$,
$\Sigma19$, $\Sigma27$, $\Sigma33$, $\Sigma41$, $\Sigma43$),
and $\langle100\rangle$ pure-twist grain boundaries ($\Sigma5$,
$\Sigma13$, $\Sigma17$, $\Sigma25$, $\Sigma29$, $\Sigma37$, $\Sigma41$).
Most of them are represented by large-scale models, hence they cannot
be calculated by DFT calculation because of the large computational
resources required. Figure \ref{fig:gbenergy} shows the rotation
angle dependence of the grain boundary energy obtained using the MLPs
and EAM potentials \citep{Ackland1987,Williams2006,Mishin1999,Zhou2004,Mishin2001}.
The values of the grain boundary energy in Al, Cu, and Pd computed
using the MLPs are consistent with those computed using the EAM potentials
and those computed by DFT calculation. Therefore, both the MLPs and
the EAM potentials have high predictive power for the grain boundary
structures and their energies. In Ag, Au, and Pt, the values of the
grain boundary energy computed using the MLPs are almost the same
as those computed by DFT calculation, whereas they deviate from those
computed using the EAM potentials. The MLPs should be more reliable
than the EAM potentials for obtaining not only the grain boundary
structures and their energies but also the other defect structures
in Ag, Au, and Pt. Note that a fine sequence is required for the component
normal to the boundary plane to obtain converged microscopic structures
when using the EAM potentials for Ag and Au. This implies that the
EAM potentials for Ag and Au lack accuracy for predicting the potential
energy surface around the globally optimal microscopic structure.

\section{Conclusion}

We have examined the predictive power of MLPs in an MLP repository
for grain boundary properties by systematically evaluating the grain
boundary energy for $\langle100\rangle$ STGBs, $\langle110\rangle$
STGBs, and $\langle100\rangle$ pure-twist grain boundaries in the
fcc elemental metals of Ag, Al, Au, Cu, Pd, and Pt. In every elemental
metal, the values of the grain boundary energy computed using the
MLP are consistent with those computed by DFT calculation. We emphasize
that the training datasets used to develop the MLPs contain no grain
boundary structures. Therefore, the consistency indicates that the
MLPs have high predictive power for the grain boundary structures
and their energies. The present results also imply that the MLPs in
the repository, including those for other systems, should be useful
in accurately predicting grain boundary properties and other complex
defect properties.
\begin{acknowledgments}
This work was supported by a Grant-in-Aid for Scientific Research
(B) (Grant Number 19H02419) and a Grant-in-Aid for Scientific Research
on Innovative Areas (Grant Number 19H05787) from the Japan Society
for the Promotion of Science (JSPS).
\end{acknowledgments}

\end{document}